# Do Environment-Modification Behaviors and Gamers' Immersiveness Shape Exceptionalism Beliefs?

Quan-Hoang Vuong [1,2], Fatemeh Kianfar [3], Thi Mai Anh Tran [4], Ni Putu Wulan Purnama Sari [5], Cresensia Dina Candra Kumaladewi [6], Viet-Phuong La [1], Minh-Hoang Nguyen [1,*]

[1] Centre for Interdisciplinary Social Research, Phenikaa University, Hanoi, Vietnam

[2] Professor, Korea University, Seoul, South Korea

[3] Faculty of Humanities, University of Hormozgan, Bandar Abbas, Hormozgan, Iran

[4] College of Forest Resources and Environmental Science, Michigan Technological University, Houghton, MI 49931 USA

[5] Faculty of Nursing, Widya Mandala Surabaya Catholic University, East Java, Indonesia

[6] Faculty of Teacher Training and Education, Widya Mandala Surabaya Catholic University, East Java, Indonesia

*Corresponding Email: hoang.nguyenminh@phenikaa-uni.edu.vn (Minh-Hoang Nguyen)

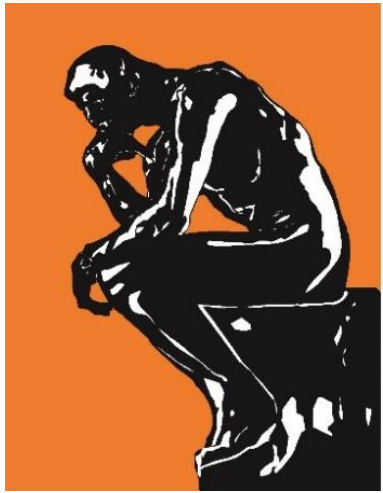

November 17, 2025

[*Original working draft v2 / Un-peer-reviewed*]

"[…] To alleviate the boredom, after catching a fish, Kingfisher would press all three buttons before swallowing the fish. Pressing the buttons has gradually become somewhat of a new technological ritual."

—In "Innovation", *Wild Wise Weird* (2024)

**Abstract**

Human exceptionalism strongly shapes human-nature perceptions, thinking, values, and behaviors. Yet little is known about how virtual ecological environments influence this mindset. As digital worlds become increasingly immersive and ecologically sophisticated, they provide novel contexts for examining how human value systems are formed and transformed. This study investigates how virtual environment-modification behaviors and players' sense of immersiveness jointly shape exceptionalism, drawing on worldviews from quantum mechanics and mathematical logic. Using Granular Interaction Thinking Theory (GITT) and the Bayesian Mindsponge Framework (BMF analytics), we analyze five key activities—tree planting, flower planting, flower crossbreeding, terraforming, and creating conditions for bug respawn—based on a multinational dataset of 640 Animal Crossing: New Horizons players from 29 countries. Results reveal two behavioral clusters distinguished by controllability. High-controllability behaviors (i.e., flower planting and terraforming) predict higher exceptionalism, whereas the flower-planting effect reverses among highly immersed players. Low-controllability behaviors (i.e., flower crossbreeding and manipulating bug spawning) predict lower exceptionalism, but these associations weaken or reverse under high immersiveness, respectively. These findings offer insights into leveraging virtual worlds to cultivate Nature Quotient (NQ), mitigate exceptionalist tendencies, and foster eco-surplus cultural orientations.

1. Introduction

Humanity is now in the Anthropocene, an era defined by our role as the major force shaping the Earth's systems (Folke et al., 2021; Steffen et al., 2011). This new and often destructive human-nature relationship (Scobie, 2023) has created escalating ecological crises rooted in deep-seated "paradigms of unsustainability" (Dewi, Winoto, Achsani, & Suprehatin, 2025; Tran et al., 2025). Central to these paradigms is anthropocentrism (Droz, 2022; Kopnina, Washington, Taylor, & J Piccolo, 2018), which manifests psychologically as Human Exceptionalism. This is a pervasive belief that humans are unique, separate from, and exempt from the planet's ecological constraints (Betz & Coley, 2022; Kim, Betz, Helmuth, & Coley, 2023). This cognitive framework, famously captured as a core component of environmental concern in the New Ecological Paradigm (Riley E. Dunlap, Van Liere, Mertig, & Jones, 2000; Xiao, Dunlap, & Hong, 2019), powerfully shapes environmental attitudes and behaviors (Kim et al., 2023).

As this global environmental context has become more urgent, a parallel digital context has become ubiquitous. Digital gaming is now a dominant global form of entertainment

(Vuong, Ho, Nguyen, et al., 2021), with "sandbox" games like Nintendo's *Animal Crossing: New Horizons* (ACNH) reaching massive audiences (e.g., 26.4 million units sold by 2021) (Fisher, Yoh, Kubo, & Rundle, 2021). These virtual worlds function as complex microcosms for human-environment interaction (Fjællingsdal & Klöckner, 2017; Vuong, Ho, Nguyen, et al., 2021). In ACNH, players are given total agency to modify their environment—terraforming land, planting or felling trees, and capturing wildlife—in a "real-time life-simulation" (Vuong et al., 2021) with few, if any, real-world consequences (Ho, Nguyen, Nguyen, La, & Vuong, 2022; Tlili, Agyemang Adarkwah, Salha, & Huang, 2024).

These activities, while ostensibly recreational, are increasingly recognized as meaningful expressions of perception and cognition, reflecting how individuals conceptualize their relationship with environments and exercise agency within artificial ecosystems (Feng, 2024; Vuong, Ho, Nguyen, et al., 2021). Consequently, these virtual environments themselves are now valued as powerful psychological tools for studying human behavior, offering high experimental control while maintaining ecological validity (Blascovich et al., 2002; Parsons, Gaggioli, & Riva, 2017; Yaremych & Persky, 2019). These games function as natural laboratories for examining ecological cognition, specifically when players are positioned as both creators and controllers, able to shape virtual ecosystems according to personal preferences with minimal constraint (Galeote, Legaki, & Hamari, 2022). This unique condition of complete human control creates an opportunity to investigate how deeply embedded worldviews, particularly anthropocentric belief systems like Exceptionalism, manifest in behavior when individuals are granted god-like powers over nature.

The potential of digital games to influence cognition and behavior has been extensively documented. Meta-analytic evidence demonstrates that action video games enhance perceptual, attentional, and cognitive skills, with measurable positive effects across domains including top-down attention and spatial cognition (Bediou et al., 2018). Beyond these cognitive dimensions, video games have shown the capacity to affect moral judgment, empathy, and environmental awareness (Schrier & Farber, 2021; Tan & Nurul-Asna, 2023). This research foundation has generated optimism about harnessing games for pro-environmental purposes. "Serious games," those designed for education, have been shown to heighten environmental awareness (Bekoum Essokolo & Robinot, 2022; J. Chen, He, Chen, & Zhang, 2024) and promote pro-environmental behaviors (Jinsong Chen, He, & She, 2025; Fox, McKnight, Sun, Maung, & Crawfis, 2020; Tan & Nurul-Asna, 2023). However, this potential is complicated by contradictory findings. Recent analyses of ACNH players reveal a troubling disconnect: players frequently engage in environmentally destructive or exploitative behaviors (e.g., unsustainable wood harvesting, exhaustive animal capture) despite holding pro-environmental perceptions (Ho et al., 2022; Vuong,

Ho, Nguyen, et al., 2021). Some players even act directly against their stated beliefs to achieve in-game goals (Tlili et al., 2024). This suggests that the simplified, consequence-free mechanics of popular games may inadvertently encourage unsustainable virtual habits.

This contradiction—that players espouse green attitudes but practice virtual exploitation—creates a significant gap in our theoretical and empirical understanding. While existing studies have documented correlations between in-game behaviors and general environmental attitudes (Ho et al., 2022; Ng, 2024; Tlili et al., 2024), the field lacks systematic investigation into the specific psychological mechanisms at work. No study to date has isolated the role of the Human Exceptionalism mindset in driving particular environment-modification behaviors within ACNH or similar virtual ecosystems. This represents a crucial oversight, as Exceptionalism constitutes a theoretically distinct cognitive construct within environmental psychology (Riley E. Dunlap et al., 2000; Kim et al., 2023), one that may operate differently from broader environmental attitudes or general pro-environmental orientation. Furthermore, a key psychological variable, immersiveness, remains unexamined. Immersiveness—the feeling of "being there" and disconnected from the real world (Michailidis, Balaguer-Ballester, & He, 2018; Weber, Weibel, & Mast, 2021)—is a powerful mechanism in virtual environments that can influence user presence and psychological outcomes (Cummings & Bailenson, 2016). It is unknown whether game immersion, by fostering a connection to a simulated nature, challenges exceptionalist thinking or, by granting players total control, reinforces it (Ahn et al., 2016; Spangenberger, Geiger, & Freytag, 2022).

To address these gaps, in this study, we employ the Granular Interaction Thinking Theory (GITT) for conceptual development and the Bayesian Mindsponge Framework (BMF) analytics for statistical analysis on a dataset of 640 ACNH players from 29 countries. We aim to examine how Human Exceptionalism manifests in virtual contexts and how immersive digital experiences modulate cognitive-behavioral relationships. Specifically, we address the following research questions:

- How are ACNH players' environment-modification behaviors in the virtual world associated with their Exceptionalism mindset?
- Do players' in-game immersiveness (i.e., perceived rich experience) moderate the association between the frequency of environment-modification behaviors and their Exceptionalism?

Our study offers several key contributions. Theoretically, the research deepens understanding of how Human Exceptionalism manifests in virtual contexts where players exert extensive environmental control, revealing how immersive digital experiences may reinforce or challenge anthropocentric worldviews. Methodologically, it demonstrates the utility of the BMF for analyzing cognitive mechanisms across culturally diverse datasets, offering a scalable and replicable model for interdisciplinary inquiry that bridges environmental psychology, game studies, and cognitive science. Practically, by isolating the operation of Exceptionalism in a simulated ecosystem, our study offers critical insight into the psychological roots of real-world unsustainable behavior. The findings provide a foundation for understanding how virtual worlds can not only illuminate but also be intentionally designed to reshape the cognitive patterns that perpetuate human-nature disconnection (Ahn, Bailenson, & Park, 2014; Markowitz & Bailenson, 2021; Santoso & Bailenson, 2020).

## 2. Methodology

### 2.1. Theoretical foundation

This study is grounded in Granular Interaction Thinking Theory (GITT), which provides a systematic framework for understanding how individuals process environmental information and translate worldviews into behavioral manifestations within virtual ecosystems (Vuong & Nguyen, 2024a, 2024b). Since the mid-twentieth century, major advances in economics have been deeply informed by concepts originating in physics, enriching our understanding of economic transactions and value formation. A prominent example is the Black–Scholes partial differential equation, whose formulation drew on Green's function techniques used in transient heat-transfer analysis—illustrating how physical principles can reshape modern socio-economic theory (Vuong, 2001). In a similar spirit, the GITT approach fosters an interdisciplinary lens that connects information processing, adaptive psychology, and human behavior.

GITT conceptualizes the human mind as a dynamic information-processing system that continuously seeks to maintain cognitive balance through selective absorption and rejection of environmental stimuli. It explains how and why people perceive and respond to environmental information based on their existing cognitive framework, with the value and impact of any experience depending on how well it aligns with the individual's internal cognitive filters (Vuong, La, & Nguyen, 2025).

According to GITT's logic, individuals continuously filter environmental information through a set of core values, absorbing inputs that align with their beliefs while ejecting those that create dissonance or challenge fundamental assumptions. This filtration process operates at a granular level, treating each discrete environmental interaction as an informational unit whose cognitive impact depends on its perceived value, contextual fit within existing belief systems, and the degree to which it reinforces or challenges pre-existing worldviews.

The translation of worldviews into behavioral expression thus involves a sequence of interrelated stages: (1) exposure to environmental information, (2) filtration through cognitive frameworks aligned with existing values, and (3) behavioral manifestation in response to perceived affordances (Vuong, La, et al., 2025).

The Animal Crossing: New Horizons (ACNH) virtual ecosystem represents an ideal environment in which to observe this process because it reproduces environmental decision-making under conditions of full player autonomy and minimal ecological or moral constraint (Blascovich et al., 2002; Parsons et al., 2017; Yaremych & Persky, 2019). Each in-game action—catching fish, felling trees, terraforming land, capturing wildlife—represents a discrete informational granule that evokes and interacts with underlying cognitive filters. Unlike physical environments, where ecological consequences create natural feedback loops that inform environmental beliefs, virtual environments offer consequence-free experimentation, making games valuable natural laboratories for examining how belief systems operate in the absence of material constraints. This granular perspective enables precise examination of how different types of virtual environmental engagement contribute to the reinforcement or transformation of environmental worldviews.

Within the GITT framework, Human Exceptionalism—the belief that humans are fundamentally separate from and exempt from ecological constraints (Betz & Coley, 2022; Riley E. Dunlap et al., 2000; Kim et al., 2023; Xu & Coley, 2022)—functions as a core value that can contribute to shaping how environmental information is processed and whether nature-exploitative behaviors are perceived as acceptable or problematic. This cognitive construct represents one component of the New Ecological Paradigm, which has been extensively validated as a measure of environmental worldviews across diverse populations and contexts (Amburgey & Thoman, 2011; Dorward et al., 2024; Riley E. Dunlap et al., 2000; Zhu & Lu, 2017). In the context of this study, a player's exceptionalism mindset functions not only as the primary filter determining which environmental stimuli are absorbed as reinforcing information and which are rejected as challenges to existing cognitive patterns, but can also be reshaped through interactions mediated by the varying levels of controllability embedded in ACNH's environment-modification activities.

In ACNH, players engage in several core environment-modification activities—planting trees, planting flowers, crossbreeding flowers, creating conditions for bug respawn, and terraforming—that differ markedly in their level of controllability. Planting trees and planting flowers are highly controllable: players directly determine the species and placement of each organism, and outcomes occur immediately and predictably. Flower crossbreeding operates under a more complex and probabilistic reproductive system, reducing controllability as players must work within genetic-like rules, adjacency requirements, and chance-based hybrid outcomes. Creating conditions for bug respawn entails even lower controllability because bug appearances depend on species-specific habitat needs, seasonal changes, weather patterns, and competition among spawn slots—factors that players can influence but cannot fully control. Terraforming provides high controllability at the structural level, allowing players to redesign the island's physical geography, but low controllability at the ecological-functional level, as players cannot dictate how species distributions, vegetation zones, and microhabitats reorganize in response to large-scale landscape changes. Through these varied interactions, players engage with varying degrees of environmental control, which may, in turn, contribute to shaping or reinforcing values in the mind, particularly regarding exceptionalism.

Immersiveness represents a critical moderating mechanism in this cognitive processing system, directly addressing our second research question (Nguyen, La, Le, & Vuong, 2022). GITT conceptualizes immersiveness not only as engagement intensity but also as a state of enhanced informational permeability—a condition where the boundaries between the player's cognitive system and the virtual environment become porous (Vuong & Nguyen, 2024b). This psychological state—the feeling of "being there" and disconnected from the real world (Michailidis et al., 2018; Weber et al., 2021)—can function as a critical moderator determining whether and how virtual environmental experiences are absorbed and internalized.

Immersive gameplay can influence information absorption and processing in various distinct directions, depending on a player's pre-existing mindset configuration and the controllability of the environment-modification activities. For players who already hold strong exceptionalism beliefs and engage in high-controllability activities, deep immersion tends to intensify the psychological experience of environmental mastery, reinforcing cognitive patterns that humans should and can dominate nature. When such players feel genuinely "present" on their virtual islands—exercising creative authority over terrain, vegetation, and species—the immersive experience transforms abstract anthropocentric beliefs into embodied experiences of dominance. The virtual world becomes an idealized arena where human-centered values operate without real-world constraints, and repeated successful manipulation of the environment strengthens exceptionalist schemas through experiential validation.

Conversely, for players with pro-environmental or eco-centric values who engage in high-controllability activities, immersion heightens their sense of connectedness to virtual nature. This intensified connection can lead to three potential outcomes. First, players' minds may gradually shift toward perceiving greater human controllability over nature, thereby weakening their initial eco-centric priors. Second, they may reduce, avoid, or engage in high-controllability actions only when necessary to achieve specific in-game goals, recognizing that excessive manipulation yields diminishing returns while consuming time and energy. Third, some players may eventually disengage from the game altogether if they perceive little benefit or increasing psychological discomfort from continuing gameplay. The sample in this study likely captures primarily the first two groups rather than the third, as individuals who stop playing are less likely to participate in ACNH-related communities where the dataset was collected.

When interactions occur through low-controllability activities, these mental outcomes shift in the opposite direction, as immersive engagement exposes players to ecological autonomy, uncertainty, and interdependence rather than mastery.

A critical implication of applying GITT to virtual environments concerns the psychological transferability of cognitive patterns cultivated through gameplay. Although virtual behaviors unfold in consequence-free settings, the underlying information-processing dynamics—particularly the repeated evaluation of human–nature relationships through either an exceptionalist or eco-centric lens—may reinforce cognitive frameworks that extend beyond the game world or, at the very least, condition the mind to become more receptive to such patterns. Virtual environments that offer thousands of micro-interactions with nature provide sustained cognitive practice in applying specific evaluative frameworks to environmental decision-making. Over time, these cumulative interactions can steer the mind toward a new cognitive equilibrium shaped by both preexisting beliefs and the novel informational inputs absorbed from the virtual environment.

Applying Bayes's theorem, such a new cognitive equilibrium can be formally expressed as follows (Vuong, Nguyen, Ho, & La, 2025).

Let $A$ denote the type of information absorbed during interactions with the virtual in-game environment. After each interaction with such information, the informational entropy state $H(X)_k$ of the mind is updated according to:

$$P(H(X)_m) = P(H(X)_k|A) = \frac{P(A|H(X)_k)P(H(X)_k)}{\sum_{i=1}^{r} P(A|H(X)_i)P(H(X)_i)}$$

Here:

- $P(H(X)_k)$ represents the prior probability distribution of informational entropy within the gamer player's mind—i.e., the cognitive state before interaction with information $A$;
- $H(X)_m$ denotes the updated entropy state after processing $A$; hence, $P(H(X)_m) = P(H(X)_k|A)$ is the posterior distribution.
- $P(A|H(X)_k)$ is the likelihood, expressing the probability that the gamer player's mind—when in a prior state $H(X)_k$—will perceive, interpret, or integrate information

$A$. This term reflects how strongly the external information resonates with, or fits into, the mind's existing cognitive–affective structure.
- The denominator $\sum_{i=1}^{r} P(A|H(X)_i)P(H(X)_i)$ represents the total probability of encountering and engaging with $A$ across all possible mental states. It acts as a normalization constant, ensuring that posterior probabilities sum to 1—effectively quantifying the overall cognitive plausibility of experiencing $A$.

Thus, the posterior distribution is proportional to:

$$P(H(X)_m) \propto P(A|H(X)_k)P(H(X)_k)$$

This relationship indicates that the gamer player's updated mental entropy state depends jointly on their prior configuration of the mind and the degree to which newly absorbed information from the virtual environment aligns with existing cognitive–affective structures. In doing so, it increases the probability that the gamer player's prior entropy state $H(X)_k$ will transition toward a new informational configuration represented by $P(H(X)_m)$.

## 2.2. Model Construction

### 2.2.1. Variable selection and generation

The study adopted an observational and cross-sectional dataset of 640 Animal Crossing: New Horizons (ACNH) players from 29 countries, which is peer-reviewed and openly available in *Data Intelligence* (Vuong, Ho, La, et al., 2021). Data collection took place between May 15 and May 30, 2020, through an online survey hosted on Google Forms and disseminated across ACNH communities on Discord, Reddit, and Facebook. Google Forms was selected for its accessibility, confidentiality, and ease of distribution via shareable links. Before circulating the survey, community administrators or moderators were contacted for approval, and the study's aims, procedures, and adherence to community regulations were fully described. The survey post—including a detailed study description and questionnaire link—was shared only after obtaining explicit permission. Participants were required to read and agree to an informed consent statement prior to participation. As an appreciation token, the first 100 respondents were given a US$5 Amazon gift card, followed by US$2 cards for the next 200 participants.

A pilot test was conducted with 15 students from Japan, Singapore, the United States, and Vietnam to ensure the clarity and reliability of the questionnaire. Their feedback informed minor adjustments to improve precision. To reduce missing data, all survey questions were set as mandatory, preventing incomplete submissions. Contact information for the research team was provided for real-time support, and all queries were addressed promptly. After data collection concluded, an official closure announcement was posted across all participating communities. Survey responses were downloaded from Google Forms in both Excel (.xls) and CSV formats. Data cleaning involved clarifying ambiguous answers, coding variables, and performing validation checks through visualization and descriptive analysis.

Among the 640 participants, 64.38% (n = 412) identified as female. The largest proportion of respondents came from the United States and Canada (55%), followed by Asia (28.13%) and the European Union (14.38%); the remaining 2.5% represented other regions. Ethnic composition reflected this geographical pattern, with White (54.22%) and Asian (31.25%) players forming the majority. Over half were undergraduate students (52.5%), and most were single and never married (61.09%). The sample largely consisted of young adults aged 18–30 (72.8%), with overall ages ranging from 11 to 55. Additionally, 63% (n = 403) reported stable employment, and only 13.44% reported not owning a pet or a garden.

For the present analysis, five variables were extracted from the dataset: one outcome variable and four predictors. The outcome variable, *Exemptionalism*, was derived from item C12 of the revised New Ecological Paradigm Scale (NEPS) developed by Riley E. Dunlap et al. (2000). This item belongs to the NEPS anti-exemptionalism subscale. It should be noted that the exceptional nature of humans was originally labeled by Catton and Dunlap (1978) as "Human Exceptionalism Paradigm," but later renamed as "Human Exemptionalism Paradigm" by Riley E Dunlap and Catton (1993), as they did not wish to deny that Homo sapiens is an "exceptional species." In this study, we use exceptionalism and excemptionalism interchangeably. Although the NEPS anti-exemptionalism sub-scale has two items, its internal consistency was relatively low (Cronbach's α = 0.49), likely due to the conceptual complexity of exemptionalism and cultural heterogeneity within the international sample. To ensure interpretive reliability, the analysis focused exclusively on item *C14*, which measures agreement with the statement: "Humans will eventually learn enough about how nature works to be able to control it" (Riley E Dunlap & Catton, 1993). Responses were recorded on a five-point Likert scale ranging from 1 ("strongly disagree") to 5 ("strongly agree").

To examine players' environment-modification behaviors in Animal Crossing: New Horizons (ACNH), respondents were asked to indicate how frequently they engaged in five key in-game ecological modification activities: planting trees, planting flowers, cross-breeding flowers, creating conditions for spawning bugs, and terraforming. These behaviors were measured on a four-point Likert scale ranging from 1 ("never") to 4 ("often"). The variables were derived from items *E15*, *E20*, *E21*, *E19*, and *E4* of the original dataset, respectively.

In addition, immersiveness was assessed using variable *F30*, which captures the extent to which players felt they were having a rich and engaging experience while playing the game.

Although this item is part of the Sensory and Imaginative Immersion subscale of the Game Experience Questionnaire—comprising six items related to narrative engagement, visual aesthetics, imagination, exploration, and overall impression—only *F30* was selected for the present analysis. This decision was made to specifically isolate and measure the players' subjective sense of immersive experience within the game.

**Table 1. Variable Description**

| Variable Name (Name in the original dataset) | Description | Data Type | Measurement |
|---|---|---|---|
| *Exemptionalism* (*C14*) | Self-reported agreement with the following statement: "Humans will eventually learn enough about how nature works to be able to control it." | Numerical | 5-point Likert scale (1 = Strongly Disagree, 5 = Strongly Agree). |
| *PlantTree* (*E15*) | Self-reported frequency of planting trees. | Numerical | 4-point Likert scale (1 = Never, 4 = Often). |
| *PlantFlower* (*E20*) | Self-reported frequency of planting flowers. | Numerical | 4-point Likert scale (1 = Never, 4 = Often). |
| *CrossbreadFlower* (*E21*) | Self-reported frequency of crossbreading flowers. | Numerical | 4-point Likert scale (1 = Never, 4 = Often). |
| *CreateBug* (*E19*) | Self-reported frequency of creating conditions for bugs to respawn. | Numerical | 4-point Likert scale (1 = Never, 4 = Often). |
| *Terraforming* (*E4*) | Self-reported frequency of terraforming. | Numerical | 4-point Likert scale (1 = Never, 4 = Often). |
| *RichExperience* (*F30*) | Self-reported feeling of rich experience while playing the game. | Numerical | 5-point Likert scale (1 = Not at all, 5 = Extremely). |

*2.2.2. Statistical Model*

To address the research question, we specified the following model to examine the relationships between players' environment-modification behaviors in the virtual world and their exceptionalism mindset, as well as the moderating role of in-game immersiveness (i.e., perceived richness of experience).

$$Exemptionalism \sim normal(\mu, \sigma) \quad (1.1)$$

$$\mu_i = \beta_0 + \beta_1 * PlantTree_i + \beta_2 * PlantFlower_i + \beta_3 * CrossbreadFlower_i + \beta_4 * CreateBug_i + \beta_5 * Terraforming_i + \beta_6 * RichExperience_i + \beta_7 * PlantTree_i * RichExperience_i + \beta_8 * PlantFlower_i * RichExperience_i + \beta_9 * CrossbreadFlower_i * RichExperience_i + \beta_{10} * CreateBug_i * RichExperience_i + \beta_{11} * Terraforming_i * RichExperience_i \quad (1.2)$$

$$\beta \sim normal(M, S) \quad (1.3)$$

The probability around $\mu$ is governed by the normal distribution, whose spread is determined by the standard deviation $\sigma$. Here, $\mu_i$ denotes game player $i$'s predicted level of exceptionalism. The predictors represent the frequency with which player $i$ engages in five types of environment-modification behaviors—planting trees, planting flowers, crossbreeding flowers, creating conditions for spawning bugs, and terraforming—as well as their perceived immersiveness while playing. The interaction terms $\beta_7$-$\beta_{11}$ capture the non-additive, moderating effects of immersiveness on the associations between each environment-modification behavior and exceptionalism. The model includes twelve parameters in total: the intercept ($\beta_0$), coefficients for the main and interaction effects ($\beta_1$-$\beta_{11}$), and the standard deviation of the residual error $\sigma$, which reflects unexplained variability in the observed data. All coefficients follow a normal prior distribution with mean $M$ and a standard deviation $S$. The logical network of the model is visualized in Figure 1.

## 2.3. Data Analysis and Validation

The present study employed the Bayesian Mindsponge Framework (BMF) analytics for several methodological and theoretical reasons. First, BMF combines the reasoning capacity of Granular Interaction Thinking Theory (GITT) with the inferential strength of Bayesian statistics, making it particularly well-suited for analyzing complex, uncertain, and information-driven psychological processes (Nguyen et al., 2022; Vuong, Nguyen, & La, 2022). Second, Bayesian inference models both known and unknown parameters as probabilistic entities (Csilléry, Blum, Gaggiotti, & François, 2010; Gill, 2014), enabling flexible and parsimonious model specification. Through Markov Chain Monte Carlo (MCMC) sampling, Bayesian methods also efficiently estimate complex structures, including multilevel and nonlinear models (Dunson, 2001). Third, compared with frequentist approaches, Bayesian analysis offers interpretative advantages—particularly the use of credible intervals, which allow continuous and probabilistic assessments of uncertainty rather than binary *p*-value–based significance testing (Halsey, Curran-Everett, Vowler, & Drummond, 2015; Wagenmakers et al., 2018).

Prior selection is central to Bayesian modeling. Given the exploratory nature of this research, uninformative (flat) priors were used to minimize prior influence on parameter estimation (Diaconis & Ylvisaker, 1985). To assess the robustness of the posterior distributions, a prior-tweaking procedure was applied by re-estimating the model with a skeptical prior representing disbelief in the associations [Normal(0, 0.5)]. If posterior estimates remain consistent across both prior settings, the results can be considered stable and insensitive to prior assumptions. The integration of priors also helps mitigate multicollinearity. As Leamer (1973) notes, multicollinearity in Bayesian inference reflects a "weak data problem," where large standard errors and the overlap of prior and posterior distributions in certain subspaces hinder inference. Empirical evidence demonstrates that models with informative priors can outperform Ridge regression in addressing multicollinearity (Adepoju & Ojo, 2018; Jaya, Tantular, & Andriyana, 2019).

Following model estimation, goodness-of-fit was evaluated using Pareto-smoothed importance sampling leave-one-out (PSIS-LOO) cross-validation (Vehtari & Gabry, 2019; Vehtari, Gelman, & Gabry, 2017). The LOO statistic is defined as:

$$LOO = -2LPPD_{loo} = -2\sum_{i=1}^{n} \log \int p(y_i|\theta)p_{post(-i)}(\theta)d\theta$$

Where $p_{post(-i)}(\theta)$ is the posterior distribution estimated after omitting observation $i$. The PSIS method uses k-Pareto values to detect influential observations. Values above 0.7 indicate highly influential cases that may distort cross-validation accuracy, whereas values below 0.5 typically reflect stable and well-fitting models.

Once the model fit was established, convergence diagnostics were conducted before interpreting the results. Convergence was assessed using both statistical and visual criteria. Statistically, the effective sample size (*n_eff*) and the Gelman–Rubin statistic (*Rhat*) were examined (Brooks & Gelman, 1998). Adequate convergence is reflected in *n_eff* >1,000 (McElreath, 2018), and *Rhat* values close to 1, with values above 1.1 indicating potential non-convergence. Visually, trace plots confirmed that the Markov chains are convergent if they mix well and exhibit stable posterior trajectories.

All Bayesian analyses were performed in R using the open-access bayesvl package version 1.0.0 (La & Vuong, 2019; Vuong & La, 2025), which provides extensive visualization tools and user-friendly model construction. To ensure transparency and reproducibility, all data and analytical scripts have been archived on Zenodo: https://zenodo.org/records/17626684

## 3. Results

The PSIS-LOO diagnostics showed that all *k*-values were below the recommended threshold of 0.5, indicating that the constructed model demonstrated a good fit to the data (Figure 2). This provides confidence that the estimated parameters can be interpreted with reliability.

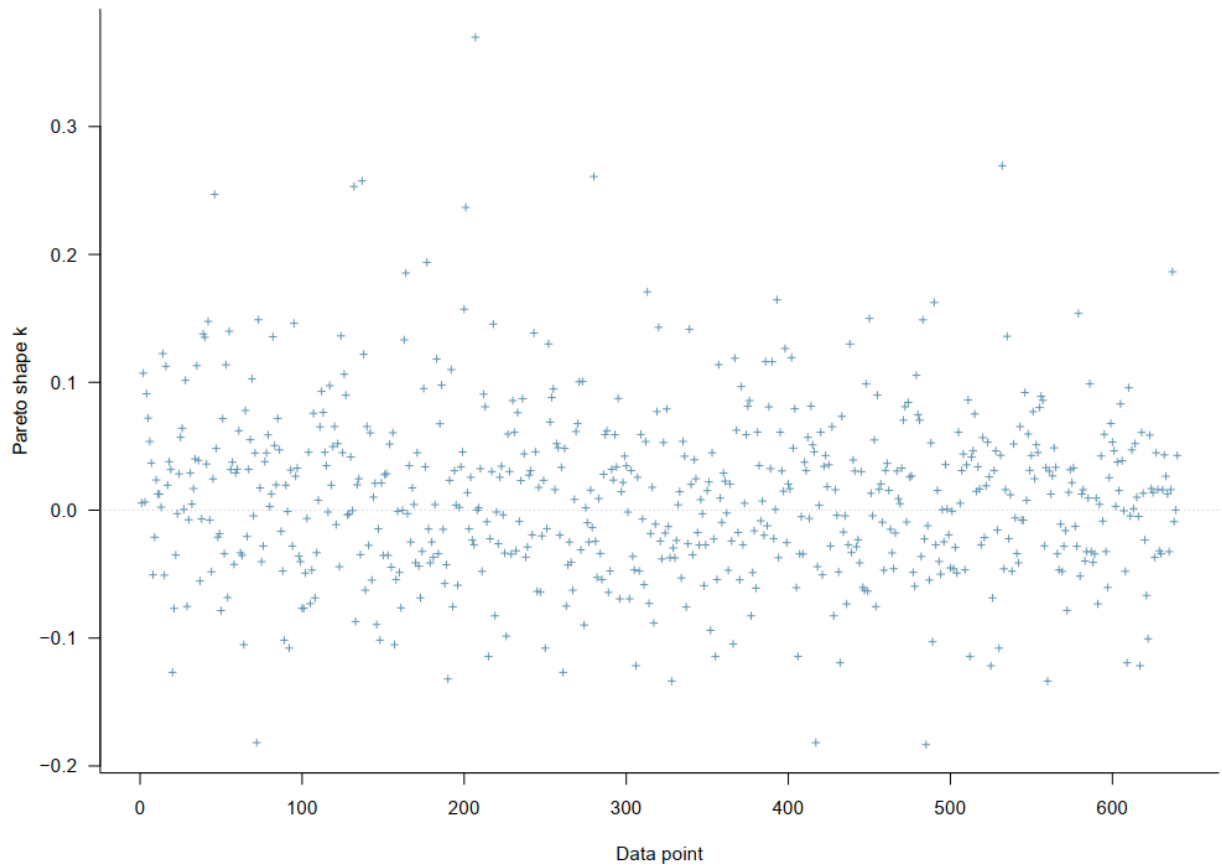

**Figure 2.** PSIS-LOO diagnostic test estimated using uninformative priors

As shown in Table 2, all posterior parameter estimates met satisfactory convergence criteria, with effective sample sizes (*n_eff*) exceeding 1,000 and *Rhat* values equal to 1. These indicators confirm that the Markov chains converged properly and provided sufficiently large effective samples for robust inference. The visual diagnostics in Figure 3 reinforce this conclusion, as the trace plots exhibit well-mixed chains that fluctuate stably around their equilibrium values.

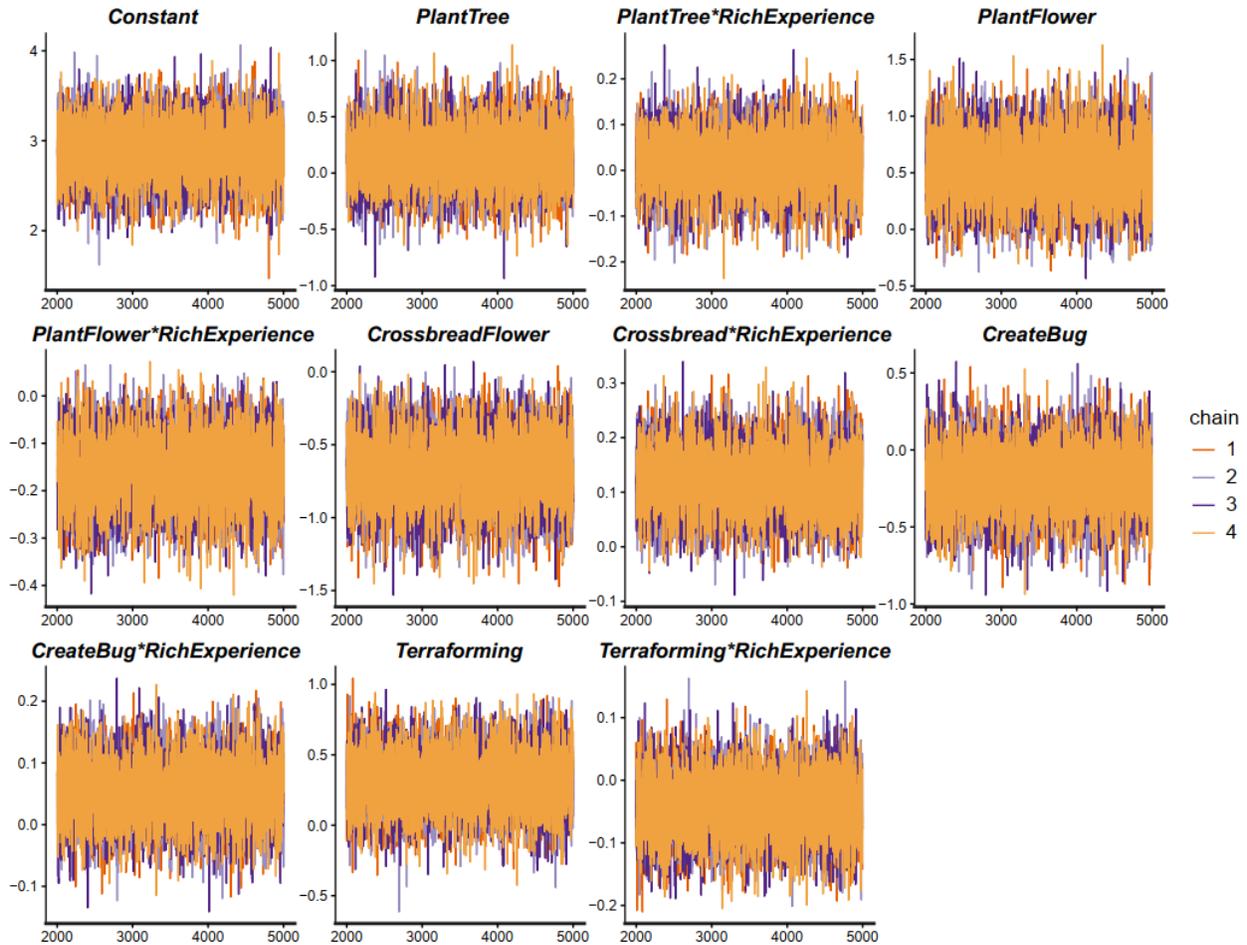

**Figure 3.** Trace plots estimated using uninformative priors

**Table 2.** Estimated posteriors

| Parameters | Uninformative priors | | | | Informative priors (Reflecting disbelief) | | | |
|---|---|---|---|---|---|---|---|---|
| | M | S | n_eff | Rhat | M | S | n_eff | Rhat |
| Constant | 2.89 | 0.30 | 11852 | 1 | 2.90 | 0.31 | 12523 | 1 |
| PlantTree | 0.15 | 0.25 | 6581 | 1 | 0.16 | 0.22 | 7451 | 1 |
| PlantTree*RichExperience | 0.02 | 0.06 | 6813 | 1 | 0.02 | 0.05 | 7682 | 1 |
| PlantFlower | 0.55 | 0.28 | 5978 | 1 | 0.37 | 0.23 | 7892 | 1 |
| PlantFlower*RichExperience | -0.16 | 0.07 | 6012 | 1 | -0.12 | 0.06 | 7346 | 1 |
| CrossbreadFlower | -0.69 | 0.22 | 6236 | 1 | -0.52 | 0.20 | 7214 | 1 |
| CrossbreadFlower*RichExperience | 0.12 | 0.05 | 6278 | 1 | 0.08 | 0.05 | 7713 | 1 |
| CreateBug | -0.20 | 0.20 | 7965 | 1 | -0.16 | 0.18 | 7962 | 1 |
| CreateBug*RichExperience | 0.05 | 0.05 | 7512 | 1 | 0.04 | 0.04 | 7854 | 1 |
| Terraforming | 0.31 | 0.20 | 7539 | 1 | 0.27 | 0.18 | 7356 | 1 |
| Terraforming*RichExperience | -0.04 | 0.05 | 7415 | 1 | -0.03 | 0.04 | 7654 | 1 |

\* Note: M = Mean, S = Standard deviation, *n_eff* = effective sample size, and *Rhat* = Gelman shrink factor

The estimated posteriors using uninformative priors are presented in Table 2, while their distributions are shown in Figure 4. In Figure 4, the blue bold lines represent the Highest Posterior Density Interval (HPDI) at 90%, the blue thin lines denote the HPDI at 99%, and the dot in the middle indicates the mean value of the distribution.

We found that the frequencies of planting flowers and terraforming are positively associated with the exceptionalism level ($M_{PlantFlower}$ = 0.55 and $S_{PlantFlower}$ = 0.28; $M_{Terraforming}$ = 0.31 and $S_{Terraforming}$ = 0.20). Both these associations are highly reliable as their HPDIs shown in Figure 4 are located entirely on the positive side of the *x*-axis. Meanwhile, the frequency of crossbreading flowers and creating conditions for bug respawn are negatively associated with the exceptionalism level ($M_{CrossbreadFlower}$ = -0.69 and $S_{CrossbreadFlower}$ = 0.22; $M_{CreateBug}$ = -0.20 and $S_{CreateBug}$ = 0.20). The 90% HDPI of $CrossbreadFlower$ is located entirely on the negative side, suggesting the coefficient's high reliability. A proportion of $CreateBug$'s HPDI is still located on the positive side, and its absolute mean value and standard deviation are equal, so its negative association can be deemed moderately reliable.

The level of immersiveness is found to positively moderate the relationship between $CrossbreadFlower$ and $Exemptionalism$ ($M_{CrossbreadFlower*RichExperience}$ = 0.12 and

$S_{CrossbreadFlower*RichExperience}$ = 0.05), and between $CreateBug$ and $Exemptionalism$ ($M_{CreateBug*RichExperience}$ = 0.05 and $S_{CreateBug*RichExperience}$ = 0.05). The former moderation can be deemed highly reliable, while the latter can only be considered moderately reliable (see Figure 4). In contrast, the level of immersiveness is found to negatively moderate the relationship between $PlantFlower$ and $Exemptionalism$ ($M_{PlantFlower*RichExperience}$ = -0.16 and $S_{PlantFlower*RichExperience}$ = 0.07).

The remaining main and interaction effects are ambiguous (e.g., $PlantTree$, $PlantTree * RichExperience$, and $Terraforming * RichExperience$). The estimated results using informative priors that reflect our disbelief in the associations are consistent with those obtained with uninformative priors. Thus, the results are deemed robust to changes in priors and to multicollinearity.

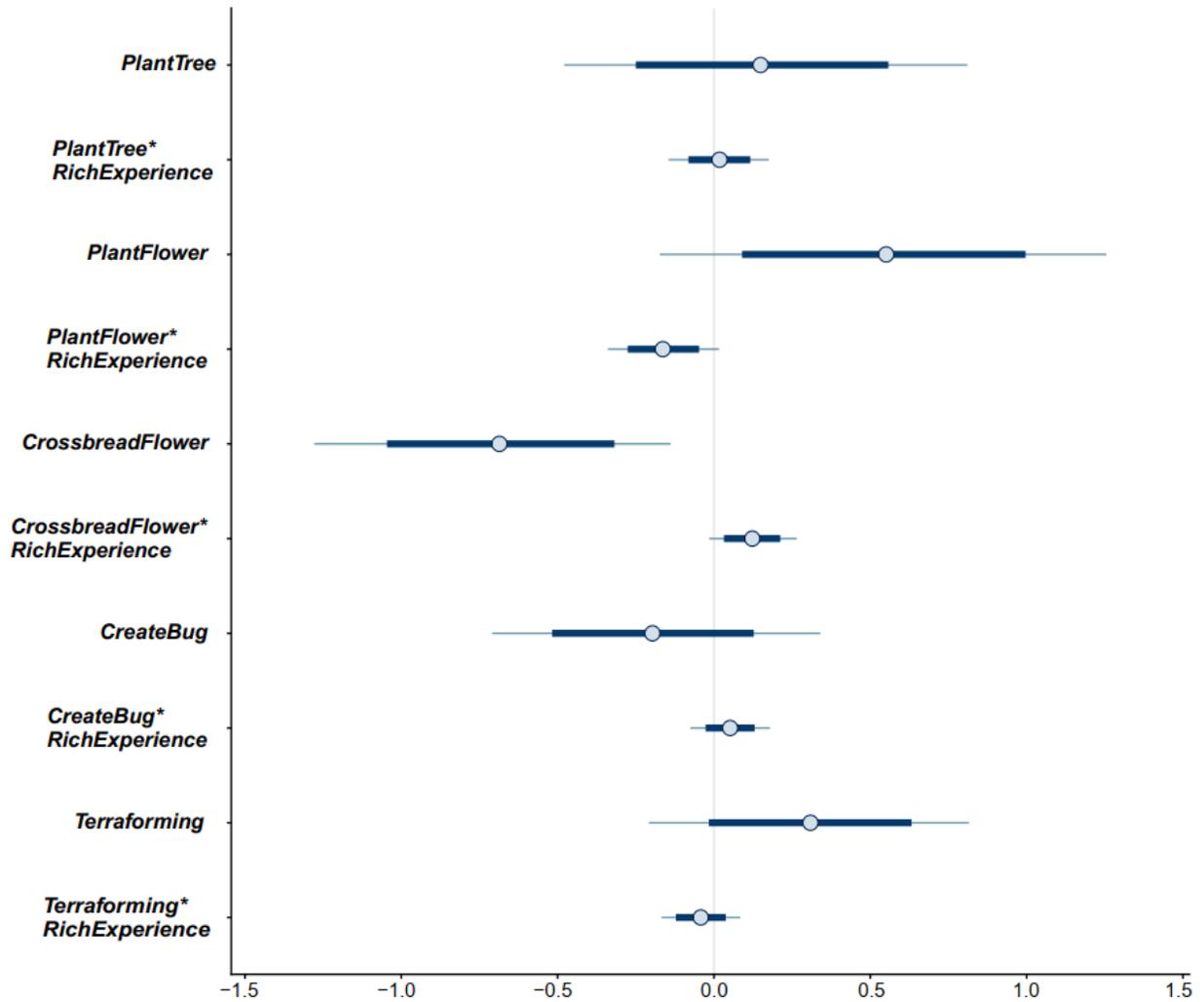

**Figure 4.** Estimated posterior distributions using uninformative priors

Employing Equation (1.2) and the posterior mean estimates from Table 2, we calculated the predicted levels of exceptionalism based on the frequency of environment-modification activities and the degree of immersiveness, focusing on the variables that exhibit clear associations with exceptionalism. The estimated exceptionalism levels for planting flowers, crossbreeding flowers, creating conditions for bug respawn, and terraforming are presented in Figures 5-a, 5-b, 5-c, and 5-d, respectively.

Figure 5-a illustrates that the association between flower-planting frequency and exceptionalism is moderated by immersiveness, proxied by the perceived richness of experience. For players who report no, slight, or moderate immersive experience, flower-planting frequency is positively associated with exceptionalism. In contrast, among those who feel fairly or extremely immersed, the association becomes negative. Additionally, among players who never plant flowers, higher immersiveness corresponds to higher exceptionalism; however, for those who sometimes or often plant flowers, higher immersiveness corresponds to lower exceptionalism.

Figure 5-b shows that the frequency of crossbreeding flowers is generally negatively associated with exceptionalism. However, the magnitude of this negative relationship depends on immersiveness: players who report a richer experience exhibit a stronger negative association. Among players who never crossbreed flowers, higher immersiveness is associated with lower exceptionalism. Conversely, among players who sometimes or often crossbreed flowers, higher immersiveness is associated with higher exceptionalism—indicating a conditional reversal depending on behavioral engagement and immersion level.

Figure 5-c indicates that the relationship between creating conditions for bug respawn and exceptionalism is also moderated by immersiveness. For players with no, slight, or moderate immersive feeling, higher bug-creation frequency is associated with lower exceptionalism. For players who feel fairly or extremely immersed, however, the association shifts to positive. A similar pattern emerges across behavioral subgroups: among players who never create bugs, higher immersiveness corresponds to lower exceptionalism, whereas among those who sometimes or often create bugs, higher immersiveness corresponds to higher exceptionalism.

Figure 5-d demonstrates that the frequency of terraforming is broadly and consistently positively associated with exceptionalism. Although higher immersiveness appears to slightly amplify this positive relationship, the magnitude of this moderating effect is statistically unreliable.

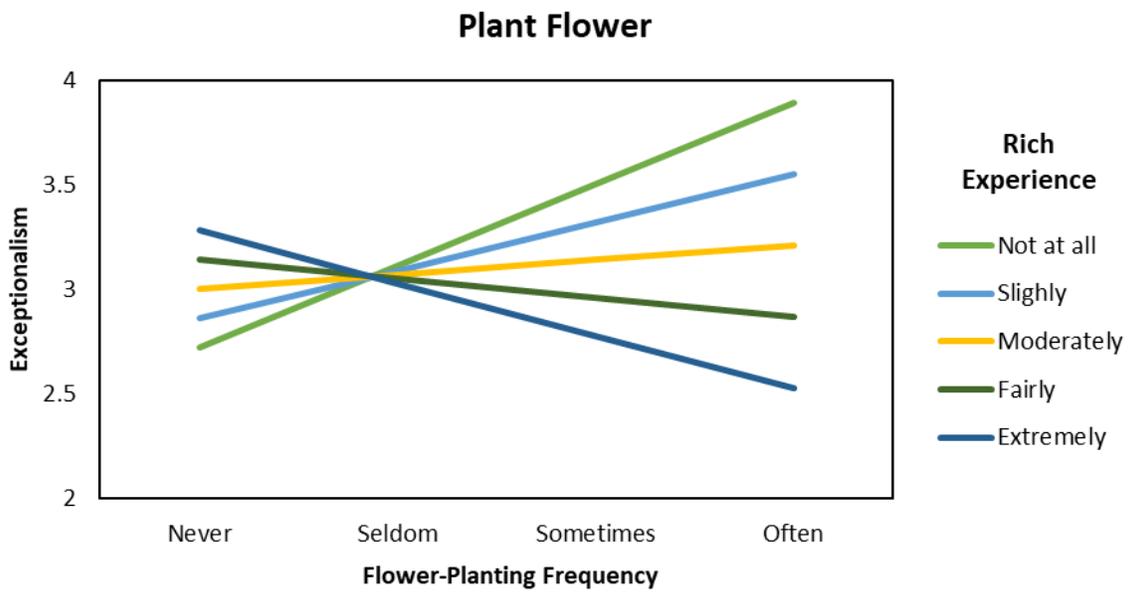

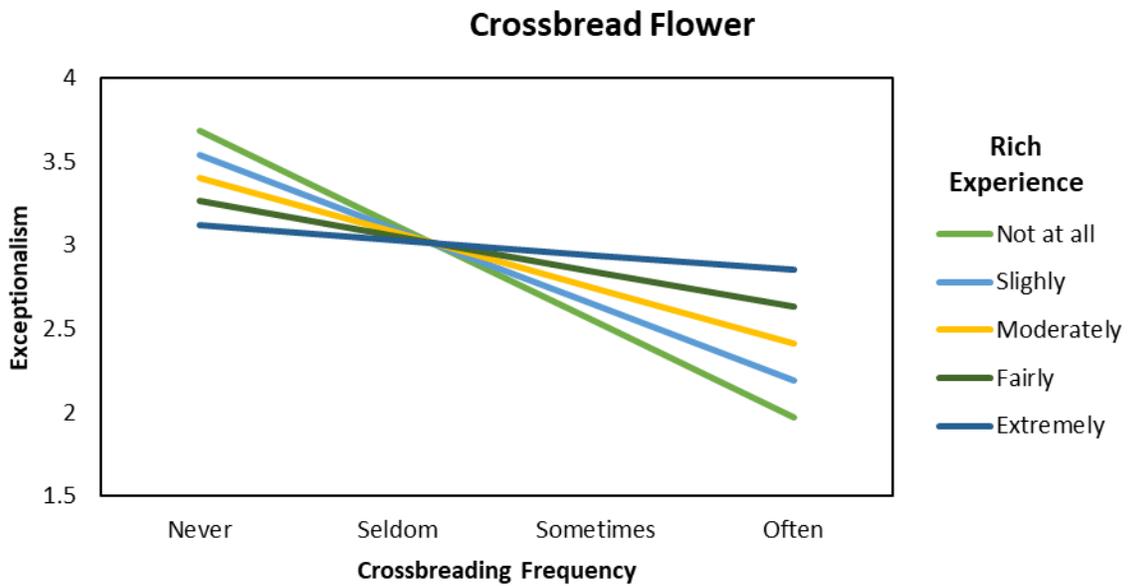

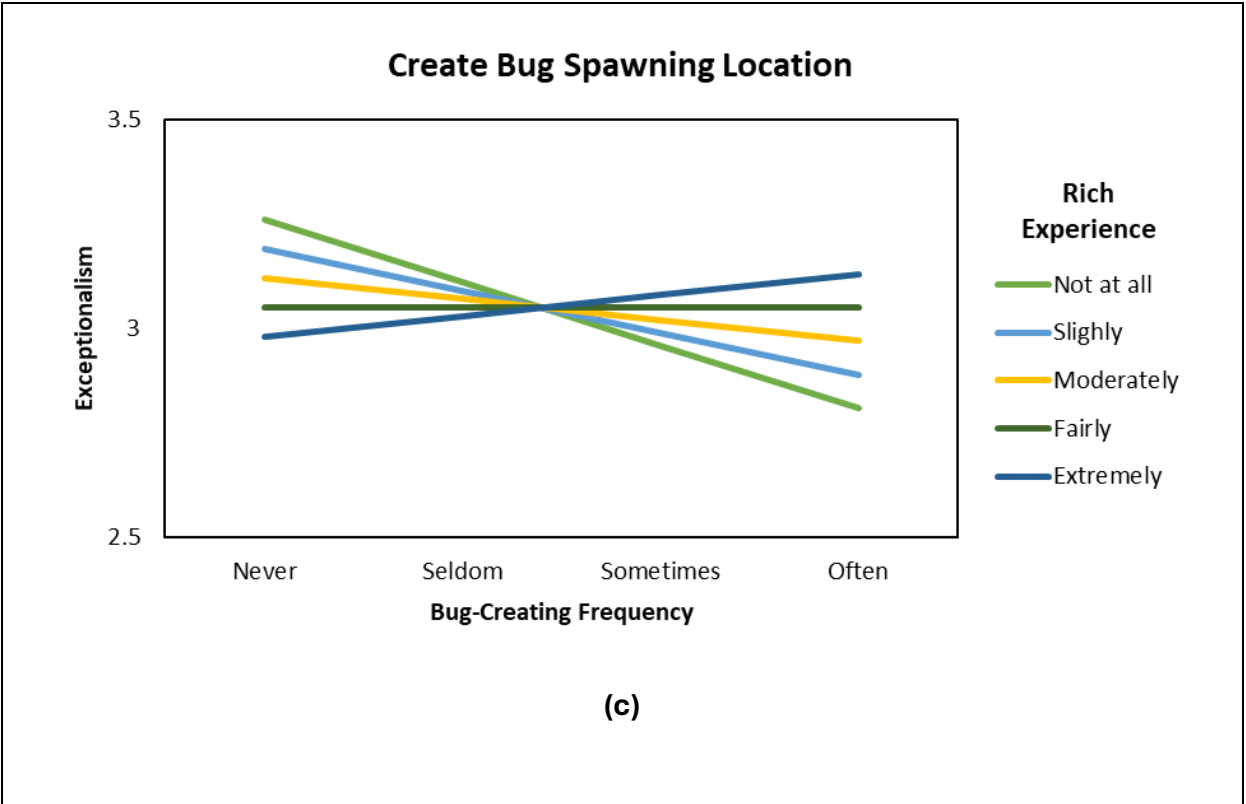

(c)

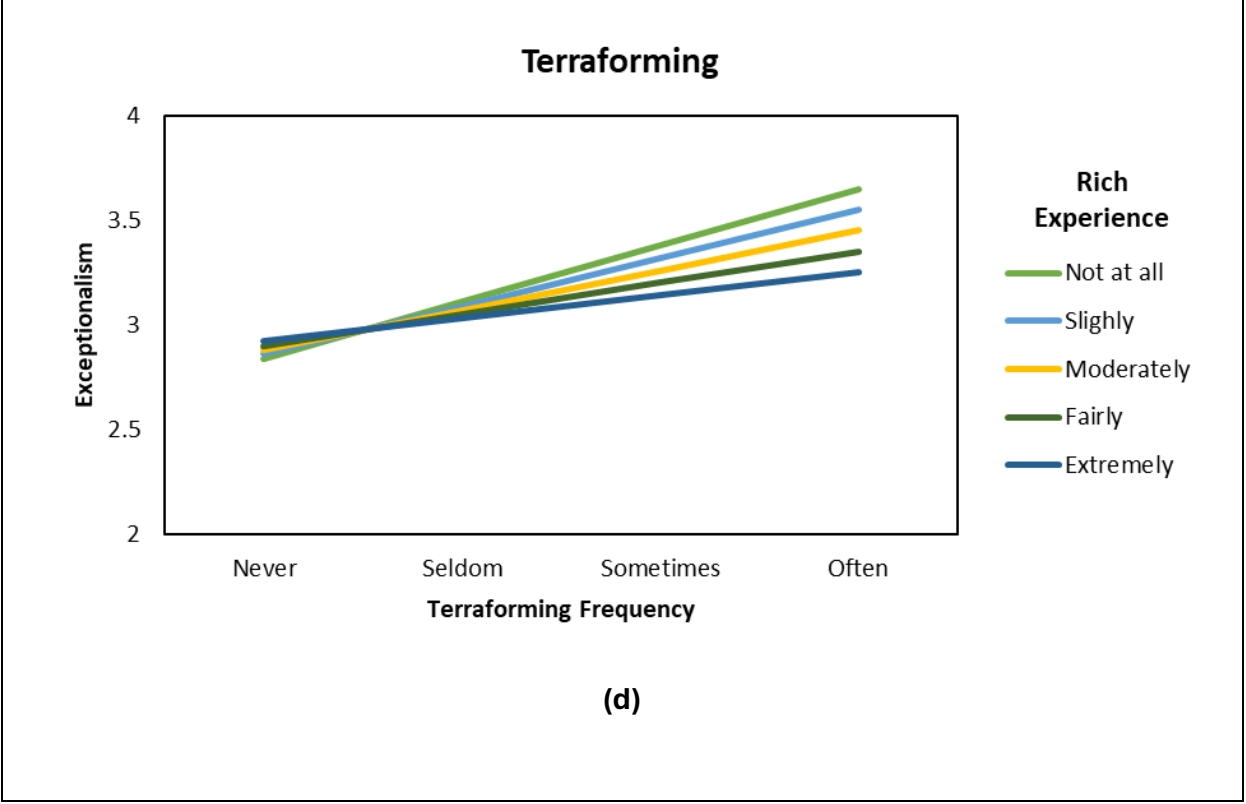

(d)

**Figure 4:** Estimated exceptionalism level of game players based on their immersiveness while playing and the frequency of a) planting flowers, b) crossbreeding flowers, c)

creating conditions for bug respawn, and d) terraforming.

## 4. Discussion

This study investigates the dynamic interplay between virtual environment-modification behaviors, players' sense of immersiveness, and human exceptionalism beliefs. Using Animal Crossing: New Horizons (ACNH) as a digital environmental context and drawing on a multinational dataset of 640 players from 29 countries, we analyze how key environment-modifying activities—tree planting, flower planting, flower crossbreeding, terraforming, and creating conditions for bug respawn—relate to players' exceptionalism tendencies and how immersiveness modifies those relationships.

The findings show that the associations between virtual environment-modifying activities and exceptionalism are complex and non-linear, varying depending on the level of immersiveness. These associations can be classified into two primary groups of activities that have different tendencies. The first group includes flower planting and terraforming activities, which allow players to have high control over the interactions with the virtual environment on their islands. Meanwhile, the second group includes flower crossbreeding and creating conditions for bug respawn, of which the interactions with the virtual environment are less controllable.

For the first group, flower-planting and terraforming frequency are positively associated with exceptionalism. However, the positive association between flower-planting frequency and exceptionalism only happens when the immersiveness is low. When the immersiveness increases, the association gradually shifts to negative. In ACNH, players only need to place flower seeds directly into the environment, and once planted, each flower becomes a static ecological element contributing to local vegetation density, habitat availability for certain bugs, and flower spreading if water. Such a process is deterministic, as the planted flower will always appear exactly as intended, with no ecological uncertainty or constraints. Meanwhile, terraforming allows players to have high control over the structural level, i.e., the island's macro-ecological landscape (e.g., constructing/demolishing cliffs, redirecting rivers, modifying coastlines, designing new land masses), which is highly difficult and complicated in reality.

Through the GITT perspective, such high-controlability activities reduce the perceived uncertainty in the player-nature interactions, shaping the perception that the players have unilateral control over the virtual ecosystem. When players do not feel deeply immersed, the informational cues absorbed from the virtual environment are relatively distant from the self, as the players do not feel "being there" (Michailidis et al., 2018; Weber et al., 2021), and thus do not form sufficient relational connection with the game content. In that circumstance, planting flowers feels like a simple, one-directional, low-uncertainty interaction with the environment, which may gradually shift the mind toward or reinforce the mindset that humans are able to control nature, i.e., exceptionalism. When players feel immersed, the mind absorbs more context-specific information from the game's ecology, fostering their relational awareness of the game's environmental constraints and feedback (e.g., growth cycles, landscape changes, spatial constraints, aesthetic ecology), as well as concerns of the game's island rating system. In ACNH, the number, distribution, and growth stage of planted flowers influence the island rating's Scenery Points. Perhaps this declined sense of control explains why highly immersive players who plant flowers more frequently are associated with lower exceptionalism. Meanwhile, terraforming does not have direct impacts on the rating system, which may explain why the association between terraforming frequency and exceptionalism is not conditional on immersiveness.

For the second group, higher frequencies of flower crossbreeding and creating conditions for bug respawn are generally associated with lower exceptionalism. However, the strength and direction of these associations differ across behaviors. Flower crossbreeding shows a consistently negative relationship with exceptionalism across levels of immersiveness, whereas the negative association for creating conditions for bug respawn is observed only among players with low immersiveness (from "not at all" to "moderately"). Among players who feel fairly or extremely immersed, the association between bug-creation frequency and exceptionalism becomes weak or even positive.

In ACNH, flower crossbreeding is governed by genetic-like hybrid rules, adjacency requirements, water dependency, and probabilistic outcomes. Creating conditions for bug respawn requires players to understand species-specific spawning rules, diurnal and seasonal cycles, habitat manipulation, and spawn-slot competition. Both activities expose players to ecological mechanisms in which they can influence environmental conditions but cannot fully determine precise or consistent results. Such low-controllability systems foster an awareness of environmental autonomy, which can reduce exceptionalist thinking.

The mechanism is particularly clear for flower crossbreeding. This activity produces dense ecological feedback that signals to players that their influence over ecological processes is limited and that nature-generated outcomes often diverge from human intention. These experiences lower exceptionalism by reinforcing humility and interdependence. The effect is strongest among players with low immersiveness, whose limited familiarity with the game may amplify perceived uncontrollability, reinforcing a non-dominant mindset. In contrast, highly immersed players—who tend to have a greater understanding and relational connection with ACNH—may perceive higher controllability even in difficult mechanics such as hybridization, thereby weakening the negative association. This reasoning also explains why, among players who never crossbreed flowers, higher immersiveness is associated with lower exceptionalism: ACNH's broader design, which emphasizes pro-environmental values and ecological appreciation, likely reinforces anti-exceptionalist orientations through other game activities (Fisher et al., 2021).

The pattern for creating conditions for bug respawn shares similarities but diverges among highly immersed players. While low-immersiveness players show a negative association between bug-creation and exceptionalism—consistent with the low-controlability ecological feedback—this relationship reverses among highly immersed players. A key contextual factor is ACNH's in-game economy: selling rare bugs is one of the most profitable activities, with high-value species (e.g., horned hercules or giraffe stag) selling for 12,000 bells, far exceeding the value of even the rarest flowers (typically around 1,000 bells). As a result, players who invest heavily in creating conditions for bug respawn often do so to maximize economic gain and other pertinent psychological (e.g., feelings of personal mastery through heightened flow and self-efficacy) and social benefits (e.g., showing status and achieving friends' admiration) (Tong et al., 2021; Yeh, Chen, Rega, & Lin, 2019). Their immersion is therefore tied to mastering economic control, learning complex spawning rules, and strategically manipulating habitats to produce lucrative outcomes. This focus on economic dominance may reinforce exceptionalist values among highly immersed players and encourage the collection of threatened species, as environmental manipulation becomes instrumentalized for financial optimization rather than ecological engagement (Fisher et al., 2021).

### 4.1. Implications

From a policy and practical standpoint, this study highlights the need for interdisciplinary collaboration among game designers, behavioral scientists, environmental educators, and public health practitioners—particularly in efforts to cultivate Nature Quotient (NQ) and foster eco-surplus cultural values (Vuong & Nguyen, 2025; Vuong, Nguyen, & La, 2025). The findings offer insights for leveraging virtual worlds as platforms for cultivating ecological knowledge, reshaping value systems, and mitigating exceptionalist cognition. Game developers can incorporate mechanics that emphasize cohabitation rather than domination—such as regenerative terraforming, collaborative habitat restoration, and non-instrumental interactions with virtual species—to provide players with experiential cues that reinforce NQ elements, including relational awareness, humility toward ecological processes, and acknowledgment of nature's autonomous agency. Landscaping and customization tools should be designed not merely for aesthetic personalization but also for ecological reflection, drawing on narrative prompts, emergent environmental behaviors, and feedback systems that help players recognize interdependence, ecological limits, and the low-controllability of outcomes inherent in non-linear ecological processes.

The study also demonstrates that the associations between environment-modification behavior frequencies and exceptionalism are moderated by immersive experience, offering additional leverage points for promoting NQ. Designers should therefore attend to how ambient soundscapes, interaction mechanics, contemplative visual environments, and emotionally resonant storytelling influence players' information processing and value formation. When games reward reciprocity, care, and ecological stewardship—rather than mastery or unilateral control—they may nudge players toward humility, relational thinking, and ecological attentiveness. Structuring immersive states around experiences of environmental reciprocity helps players internalize ecological autonomy and the non-linear dynamics of ecosystems, thereby strengthening the cognitive foundations of eco-surplus culture.

Because virtual acts such as planting trees, crossbreeding flowers, terraforming, or manipulating bug spawning serve as meaningful proxies for real-world environmental thinking, researchers and educators can use these environments as scalable laboratories for studying cognitive processing, value change, and NQ development. Behavioral data emerging from virtual platforms can inform interventions designed to weaken exceptionalist values and enhance nature-related cognitive flexibility, particularly among youth and digital-native populations. Environmental education programs can integrate sandbox games like ACNH to guide learners in exploring ecological functions, interactions, and systems thinking, thereby reinforcing ecological humility and reflective environmental identity.

Finally, the findings suggest that digital ecologies should be recognized as useful spaces for fostering NQ and eco-surplus culture, especially in urban contexts or situations where physical access to nature is limited. Policymakers can support initiatives that use virtual platforms to promote pro-environmental values, strengthen ecological empathy, and cultivate psychological connectedness to nature within sustainability campaigns, climate education efforts, mental health interventions, and civic engagement strategies. By acknowledging the influence of virtual nature on human value systems—and by embedding NQ-enhancing features into virtual design—policymakers and educators can expand the societal toolkit for environmental value transformation in an increasingly digital era.

### 4.2. Study Limitations

Despite its novel contributions to understanding exceptionalism through virtual environment-modification behaviors, this study presents several limitations that warrant consideration. First, the use of self-reported data may introduce interpretive ambiguity or response bias. Second, the nature of the cross-sectional design limits causal inference. While associations between virtual environmental modification behaviors and exceptionalism are robust, certain temporal dynamics, such as how exceptionalist beliefs evolve with gameplay, remain unexplored. Third, although the sample spans 29 countries, cultural interpretations of nature, gaming, and symbolic agency may vary significantly. Exceptionalism may manifest differently across sociocultural contexts, potentially affecting the generalizability of findings. Fourth, this study interprets virtual environmental modification behaviors as symbolic acts, yet these may also be driven by gameplay optimization or aesthetic preferences unrelated to exceptionalist cognition. Disentangling symbolic agency from ludic functionality remains a methodological challenge, especially in sandbox-style games like ACNH. Fifth, while rich experience is treated as a moderator, its operationalization may not fully capture the depth and nuance of immersive states. Lastly, this study focuses exclusively on ACNH gaming, which has unique/specific mechanics and cultural aesthetics. Findings may not translate to other virtual environments with different affordances, such as survival games (e.g., Minecraft) or competitive platforms (e.g., EVE Online).

**References**

Adepoju, A. A., & Ojo, O. O. (2018). Bayesian method for solving the problem of multicollinearity in regression. *Afrika Statistika, 13*(3), 1823-1834. doi:10.16929/as/1823.135

Ahn, S. J., Bailenson, J. N., & Park, D. (2014). Short- and long-term effects of embodied experiences in immersive virtual environments on environmental locus of control and behavior. *Computers in Human Behavior, 39*, 235-245. doi:https://doi.org/10.1016/j.chb.2014.07.025

Ahn, S. J., Bostick, J., Ogle, E., Nowak, K. L., McGillicuddy, K. T., & Bailenson, J. N. (2016). Experiencing Nature: Embodying Animals in Immersive Virtual Environments Increases Inclusion of Nature in Self and Involvement with Nature. *Journal of Computer-Mediated Communication, 21*(6), 399-419. doi:10.1111/jcc4.12173

Amburgey, J. W., & Thoman, D. B. (2011). Dimensionality of the New Ecological Paradigm: Issues of Factor Structure and Measurement. *Environment and Behavior, 44*(2), 235-256. doi:10.1177/0013916511402064

Bediou, B., Adams, D. M., Mayer, R. E., Tipton, E., Green, C. S., & Bavelier, D. (2018). Meta-analysis of action video game impact on perceptual, attentional, and cognitive skills. *Psychol Bull, 144*(1), 77-110. doi:10.1037/bul0000130

Bekoum Essokolo, V.-L., & Robinot, E. (2022). «Let's Go Deep into the Game to Save Our Planet!» How an Immersive and Educational Video Game Reduces Psychological Distance and Raises Awareness. *Sustainability, 14*(10). doi:10.3390/su14105774

Betz, N., & Coley, J. D. (2022). Human Exceptionalist Thinking about Climate Change. *Sustainability, 14*(15). doi:10.3390/su14159519

Blascovich, J., Loomis, J., Beall, A. C., Swinth, K. R., Hoyt, C. L., & Bailenson, J. N. (2002). Immersive Virtual Environment Technology as a Methodological Tool for Social Psychology. *Psychological Inquiry, 13*(2), 103-124.
Brooks, S. P., & Gelman, A. (1998). General methods for monitoring convergence of iterative simulations. *Journal of computational and graphical statistics, 7*(4), 434-455.
Catton, W. R., & Dunlap, R. E. (1978). Environmental sociology: A new paradigm. *The American Sociologist, 31*(1), 41-49.
Chen, J., He, M., Chen, J., & Zhang, C. (2024). Exploring the role and mechanisms of environmental serious games in promoting pro-environmental decision-making: a focused literature review and future research agenda. *Front Psychol, 15*, 1455005. doi:10.3389/fpsyg.2024.1455005
Chen, J., He, M., & She, S. (2025). The impact of environmental serious game on pro-environmental behavior through environmental psychological ownership and environmental self-efficacy. *Scientific Reports, 15*(1), 27616. doi:10.1038/s41598-025-11297-z
Csilléry, K., Blum, M. G., Gaggiotti, O. E., & François, O. (2010). Approximate Bayesian computation (ABC) in practice. *Trends in Ecology and Evolution, 25*(7), 410-418. doi:10.1016/j.tree.2010.04.001
Cummings, J. J., & Bailenson, J. N. (2016). How Immersive Is Enough? A Meta-Analysis of the Effect of Immersive Technology on User Presence. *Media Psychology, 19*(2), 272-309. doi:10.1080/15213269.2015.1015740
Dewi, D. E., Winoto, J., Achsani, N. A., & Suprehatin, S. (2025). Understanding Deep-Seated Paradigms of Unsustainability to Address Global Challenges: A Pathway to Transformative Education for Sustainability. *World, 6*(3). doi:10.3390/world6030106
Diaconis, P., & Ylvisaker, D. (1985). Quantifying prior opinion. In J. M. Bernardo, M. H. DeGroot, D. V. Lindley, & A. F. M. Smith (Eds.), *Bayesian statistics* (Vol. 2, pp. 133-156): North Holland Press.
Dorward, L., Ibbett, H., Dwiyahreni, A. A., Kohi, E., Prayitno, K., Sankeni, S., ... St John, F. A. V. (2024). Cross-Cultural Applications of the New Ecological Paradigm in Protected Area Contexts. *Environ Behav, 56*(1-2), 120-151. doi:10.1177/00139165241274623
Droz, L. (2022). Anthropocentrism as the scapegoat of the environmental crisis: a review. *Ethics in Science and Environmental Politics, 22*, 25-49.
Dunlap, R. E., & Catton, W. R. (1993). The development, current status, and probable future of environmental sociology: Toward an ecological sociology. *Annals of the International Institute of Sociology, 3*, 263-284.
Dunlap, R. E., Van Liere, K. D., Mertig, A. G., & Jones, R. E. (2000). New Trends in Measuring Environmental Attitudes: Measuring Endorsement of the New Ecological Paradigm: A Revised NEP Scale. *Journal of Social Issues, 56*(3), 425-442. doi:https://doi.org/10.1111/0022-4537.00176
Dunson, D. B. (2001). Commentary: practical advantages of Bayesian analysis of epidemiologic data. *American Journal of Epidemiology, 153*(12), 1222-1226. doi:10.1093/aje/153.12.1222

Feng, Z. (2024). How Game Ecology Drives Player Choices in Strategy Video Games: An Environmental Determinism Perspective. *Simulation & Gaming, 55*(2), 342-357. doi:10.1177/10468781241231895

Fisher, J. C., Yoh, N., Kubo, T., & Rundle, D. (2021). Could Nintendo's Animal Crossing be a tool for conservation messaging? *People and Nature, 3*(6), 1218-1228. doi:https://doi.org/10.1002/pan3.10240

Fjællingsdal, K. S., & Klöckner, C. A. (2017). ENED-GEM: A Conceptual Framework Model for Psychological Enjoyment Factors and Learning Mechanisms in Educational Games about the Environment. *Frontiers in Psychology, Volume 8 - 2017*. doi:10.3389/fpsyg.2017.01085

Folke, C., Polasky, S., Rockström, J., Galaz, V., Westley, F., Lamont, M., . . . Walker, B. H. (2021). Our future in the Anthropocene biosphere. *Ambio, 50*(4), 834-869. doi:10.1007/s13280-021-01544-8

Fox, J., McKnight, J., Sun, Y., Maung, D., & Crawfis, R. (2020). Using a serious game to communicate risk and minimize psychological distance regarding environmental pollution. *Telematics and Informatics, 46*, 101320. doi:https://doi.org/10.1016/j.tele.2019.101320

Galeote, D. F., Legaki, N.-Z., & Hamari, J. (2022). *Avatar Identities and Climate Change Action in Video Games: Analysis of Mitigation and Adaptation Practices*. Paper presented at the Proceedings of the 2022 CHI Conference on Human Factors in Computing Systems, New Orleans, LA, USA. https://doi.org/10.1145/3491102.3517438

Gill, J. (2014). *Bayesian methods: A social and behavioral sciences approach* (Vol. 20): CRC Press.

Halsey, L. G., Curran-Everett, D., Vowler, S. L., & Drummond, G. B. (2015). The fickle P value generates irreproducible results. *Nature Methods, 12*, 179-185. doi:10.1038/nmeth.3288

Ho, M. T., Nguyen, T. T., Nguyen, M. H., La, V. P., & Vuong, Q. H. (2022). Good ethics cannot stop me from exploiting: The good and bad of anthropocentric attitudes in a game environment. *Ambio, 51*(11), 2294-2307. doi:10.1007/s13280-022-01742-y

Jaya, I., Tantular, B., & Andriyana, Y. (2019). *A Bayesian approach on multicollinearity problem with an Informative Prior*. Paper presented at the Journal of Physics: Conference Series.

Kim, J. J. H., Betz, N., Helmuth, B., & Coley, J. D. (2023). Conceptualizing Human–Nature Relationships: Implications of Human Exceptionalist Thinking for Sustainability and Conservation. *Topics in Cognitive Science, 15*(3), 357-387. doi:https://doi.org/10.1111/tops.12653

Kopnina, H., Washington, H., Taylor, B., & J Piccolo, J. (2018). Anthropocentrism: More than Just a Misunderstood Problem. *Journal of Agricultural and Environmental Ethics, 31*(1), 109-127. doi:10.1007/s10806-018-9711-1

La, V.-P., & Vuong, Q.-H. (2019). bayesvl: Visually learning the graphical structure of Bayesian networks and performing MCMC with 'Stan'. *The Comprehensive R Archive Network (CRAN)*.

Leamer, E. E. (1973). Multicollinearity: a Bayesian interpretation. *The Review of Economics and Statistics, 55*(3), 371-380.
Markowitz, D. M., & Bailenson, J. N. (2021). Virtual reality and the psychology of climate change. *Current Opinion in Psychology, 42*, 60-65. doi:https://doi.org/10.1016/j.copsyc.2021.03.009
McElreath, R. (2018). *Statistical rethinking: A Bayesian course with examples in R and Stan*: Chapman and Hall/CRC.
Michailidis, L., Balaguer-Ballester, E., & He, X. (2018). Flow and Immersion in Video Games: The Aftermath of a Conceptual Challenge. *Frontiers in Psychology, Volume 9 - 2018*. doi:10.3389/fpsyg.2018.01682
Ng, Y.-L. (2024). Uses and Gratifications of Biophilic Simulation Games. *Games and Culture*, 15554120241249518. doi:10.1177/15554120241249518
Nguyen, M.-H., La, V.-P., Le, T.-T., & Vuong, Q.-H. (2022). Introduction to Bayesian Mindsponge Framework analytics: An innovative method for social and psychological research. *MethodsX, 9*, 101808. doi:10.1016/j.mex.2022.101808
Parsons, T. D., Gaggioli, A., & Riva, G. (2017). Virtual Reality for Research in Social Neuroscience. *Brain Sci, 7*(4). doi:10.3390/brainsci7040042
Santoso, M., & Bailenson, J. (2020). Virtual Reality Experiences to Promote Environmental Climate Citizenship. In M. Lackner, B. Sajjadi, & W.-Y. Chen (Eds.), *Handbook of Climate Change Mitigation and Adaptation* (pp. 1-43). New York, NY: Springer New York.
Schrier, K., & Farber, M. (2021). A systematic literature review of 'empathy' and 'games'. *Journal of Gaming & Virtual Worlds, 13*(2), 195-214. doi:https://doi.org/10.1386/jgvw_00036_1
Scobie, M. (2023). Framing Environmental Human Rights in the Anthropocene. In W. F. Baber & J. R. May (Eds.), *Environmental Human Rights in the Anthropocene: Concepts, Contexts, and Challenges* (pp. 9-30). Cambridge: Cambridge University Press.
Spangenberger, P., Geiger, S. M., & Freytag, S.-C. (2022). Becoming nature: effects of embodying a tree in immersive virtual reality on nature relatedness. *Scientific Reports, 12*(1), 1311. doi:10.1038/s41598-022-05184-0
Steffen, W., Persson, A., Deutsch, L., Zalasiewicz, J., Williams, M., Richardson, K., . . . Svedin, U. (2011). The anthropocene: from global change to planetary stewardship. *Ambio, 40*(7), 739-761. doi:10.1007/s13280-011-0185-x
Tan, C. K. W., & Nurul-Asna, H. (2023). Serious games for environmental education. *Integrative Conservation, 2*(1), 19-42. doi:https://doi.org/10.1002/inc3.18
Tlili, A., Agyemang Adarkwah, M., Salha, S., & Huang, R. (2024). How environmental perception affects players' in-game behaviors? Towards developing games in compliance with sustainable development goals. *Entertainment Computing, 50*, 100678. doi:https://doi.org/10.1016/j.entcom.2024.100678
Tong, X., Gromala, D., Neustaedter, C., Fracchia, F. D., Dai, Y., & Lu, Z. (2021). Players' Stories and Secrets in Animal Crossing: New Horizons-Exploring Design Factors for Positive Emotions and Social Interactions in a Multiplayer Online Game. *Proceedings of the ACM on Human-Computer Interaction, 5*(CHI PLAY), 1-23.

Tran, T. M. A., Reed-VanDam, C., Belopavlovich, K., Brown, E., Higdon, K., Lane-Clark, S. N., . . . Gagnon, V. (2025). Decentering humans in sustainability: a framework for Earth-centered kinship and practice. *Socio-Ecological Practice Research, 7*(1), 43-55. doi:10.1007/s42532-024-00206-9

Vehtari, A., & Gabry, J. (2019). Bayesian Stacking and Pseudo-BMA weights using the loo package (Version loo 2.2.0). Retrieved from https://mc-stan.org/loo/articles/loo2-weights.html

Vehtari, A., Gelman, A., & Gabry, J. (2017). Practical Bayesian model evaluation using leave-one-out cross-validation and WAIC. *Statistics and computing, 27*(5), 1413-1432. doi:10.1007/s11222-016-9696-4

Vuong, Q.-H. (2001). Black-Scholes PDE: A finance application. In Vietnam National University and Institute of Mathematics (Ed.), *Proceedings of the International Conference on Differential Equation Approximation and Applications*: Vietnam National University and Institute of Mathematics.

Vuong, Q.-H. (2024). *Wild Wise Weird*: AISDL.

Vuong, Q.-H., Ho, M.-T., La, V.-P., Le, T.-T., Nguyen, T. H. T., & Nguyen, M.-H. (2021). A multinational data set of game players' behaviors in a virtual world and environmental perceptions. *Data Intelligence, 3*(4), 606-630. doi:10.1162/dint_a_00111

Vuong, Q.-H., Ho, M.-T., Nguyen, M.-H., Pham, T.-H., Vuong, T.-T., Khuc, Q., ... La, V.-P. (2021). On the environment-destructive probabilistic trends: A perceptual and behavioral study on video game players. *Technology in Society, 65*, 101530. doi:https://doi.org/10.1016/j.techsoc.2021.101530

Vuong, Q.-H., & La, V.-P. (2025). *Package 'bayesvl' version 1.0.0*.

Vuong, Q.-H., La, V.-P., & Nguyen, M.-H. (2025). Informational entropy-based value formation: A new paradigm for a deeper understanding of value. *Evaluation Review*. doi:10.1177/0193841X251396210

Vuong, Q.-H., & Nguyen, M.-H. (2024a). *Better economics for the Earth: A lesson from quantum and information theories*: AISDL.

Vuong, Q.-H., & Nguyen, M.-H. (2024b). Exploring the role of rejection in scholarly knowledge production: Insights from granular interaction thinking and information theory. *Learned Publishing, 37*(4), e1636. doi:10.1002/leap.1636

Vuong, Q.-H., & Nguyen, M.-H. (2025). On Nature Quotient. *Pacific Conservation Biology, 31*, PC25028. doi:10.1071/PC25028

Vuong, Q.-H., Nguyen, M.-H., Ho, M.-T., & La, V.-P. (2025). *GITT: Essentials, Uses, and Usage*: AISDL.

Vuong, Q.-H., Nguyen, M.-H., & La, V.-P. (2022). *The mindsponge and BMF analytics for innovative thinking in social sciences and humanities*: Walter de Gruyter GmbH.

Vuong, Q.-H., Nguyen, M.-H., & La, V.-P. (2025). Towards an eco-surplus culture. From market failures to vulnerabilities and logical flaws of 'artificial' environmental protection systems. *Visions for Sustainability, 24*(12178), 1-68. doi:10.13135/2384-8677/12178

Wagenmakers, E.-J., Marsman, M., Jamil, T., Ly, A., Verhagen, J., Love, J., ... Epskamp, S. (2018). Bayesian inference for psychology. Part I: Theoretical advantages and

practical ramifications. *Psychonomic Bulletin Review, 25*(1), 35-57. doi:10.3758/s13423-017-1343-3

Weber, S., Weibel, D., & Mast, F. W. (2021). How to Get There When You Are There Already? Defining Presence in Virtual Reality and the Importance of Perceived Realism. *Frontiers in Psychology, Volume 12 - 2021*. doi:10.3389/fpsyg.2021.628298

Xiao, C., Dunlap, R. E., & Hong, D. (2019). Ecological Worldview as the Central Component of Environmental Concern: Clarifying the Role of the NEP. *Society & Natural Resources, 32*(1), 53-72. doi:10.1080/08941920.2018.1501529

Xu, Y., & Coley, J. D. (2022). Intuitive biological thinking in Chinese 8th graders. *Journal of Experimental Child Psychology, 224*, 105511. doi:https://doi.org/10.1016/j.jecp.2022.105511

Yaremych, H. E., & Persky, S. (2019). Tracing physical behavior in virtual reality: A narrative review of applications to social psychology. *Journal of Experimental Social Psychology, 85*, 103845. doi:https://doi.org/10.1016/j.jesp.2019.103845

Yeh, Y.-c., Chen, S.-Y., Rega, E. M., & Lin, C.-S. (2019). Mindful learning experience facilitates mastery experience through heightened flow and self-efficacy in game-based creativity learning. *Frontiers in psychology, 10*, 1593.

Zhu, X., & Lu, C. (2017). Re-evaluation of the New Ecological Paradigm scale using item response theory. *Journal of Environmental Psychology, 54*, 79-90. doi:https://doi.org/10.1016/j.jenvp.2017.10.005